\documentstyle[twoside,fleqn,espcrc2,epsf]{article}
\newcommand{\be}{\begin{equation}}
\newcommand{\ee}{\end{equation}}
\newcommand{\bea}{\begin{eqnarray}}
\newcommand{\eea}{\end{eqnarray}}
\newcommand{\bean}{\begin{eqnarray*}}
\newcommand{\eean}{\end{eqnarray*}}

\newcommand{\AmS}{{\protect\the\textfont2
  A\kern-.1667em\lower.5ex\hbox{M}\kern-.125emS}}

\title{Hybrid Quarkonia with High Statistics from NRQCD}

\author{UKQCD Collaboration - T. Manke, I.T. Drummond, R.R. Horgan,
        H.P. Shanahan 
        \address{DAMTP,  Silver Street, Cambridge CB3 9EW, ENGLAND}%
        \thanks{Talk presented by T. Manke. Supported in part by EU grant ERBCHB6-CT94-0523 and
        EPSRC 94007885.
        Our calculations were performed at the HPCF
        at Cambridge University.}
}
       
\begin{document}

\begin{abstract}
We have studied the $O(mv^6)$ effects in NRQCD on the
spectrum of heavy quarkonia and compare our results for different
lattices (quenched and dynamical). We also report on an investigation
into hybrid states within the framework of NRQCD. This suggests
that the lowest lying hybrid is around the $B^* \bar B$ threshold and 
3 standard deviations above the $B \bar B$.
\end{abstract}
%
\maketitle
\section{INTRODUCTION}
NRQCD has proved to be very successful when applied to heavy quark spectroscopy
as relativistic effects can be taken into account systematically.
In particular the low lying experimental spectrum can be reproduced
directly from ${\cal L}_{\rm NRQCD}$. Because NRQCD allows very fast
and accurate measurements it is also possible to obtain reliable
results for the spin structure which reveals the quark content in the
confining regime. In a previous paper \cite{omv6} we reported on a
study of higher order corrections to the spin splittings. Here we will
analyse these results more carefully on different lattices including
some preliminary studies on unquenched configurations.
It is very appealing to reveal also the gluon content in the
non-perturbative regime of QCD. To this end we couple the magnetic
field to the quark-antiquark pair. This hybrid state will allow quantum
numbers which are not permitted in the ordinary quark model.
There are many predictions for the masses of such hybrids from phenomenological
models \cite{bag,fluxtube} and from lattice calculations in
the static limit \cite{cmhybrid}.
Here we go beyond the static approximation and employ
NRQCD to treat the quarks dynamically.
In Section 2 we present our Lagrangian and the results from
a detailed analysis of the systematic errors in NRQCD.
In Section 3 we report on hybrid results from NRQCD.
\section{SYSTEMATIC ERRORS}
For our investigation into the systematic errors of NRQCD we
use the evolution equation
\bea
G_{t+1} &=& K_{t+1}~U_t^{\dag}(x)~K_t~\left(1-a\delta H \right) G_{t\ge
1} ~~,\nonumber \\ 
G_{t=1} &=&  K_{t=1}~U_{t=0}^{\dag}(x)~K_{t=0}~S({\bf x})~~, \nonumber \\
K_t &=& \left(1-\frac{aH_0}{2n}\right)_t^n~~,~~H_0 = -
\frac{\Delta^2}{2m_b}~~.
\label{eq:evol}
\eea
The source term, $S({\bf x})$, depends on the operator
to be propagated and $\delta H$ is given in \cite{omv6} including terms
up to $O(mv^6)$. The stability parameter $n$ depends
on the quark mass $m_b$.
We also reduced the lattice artefacts
by redefining the lattice derivatives and the
electromagnetic field tensor in Equation \ref{eq:evol}. 
All links are tadpole improved and we use the tree-level coefficients
for all operators. For our analysis with lower accuracy in the
velocity expansion we use the same evolution equation as in
\cite{cdomv4}. As opposed to \cite{cdomv4} we employed gauge invariant
operators throughout. We constructed the meson operators with definite
$J^{PC}$ from extended link variables to improve the overlap with the
ground state \cite{omv6}.
\subsection{$O(mv^4)$ vs. $O(mv^6)$}
\begin{figure}[htb]
    \hbox{\epsfxsize = 72mm  \epsffile{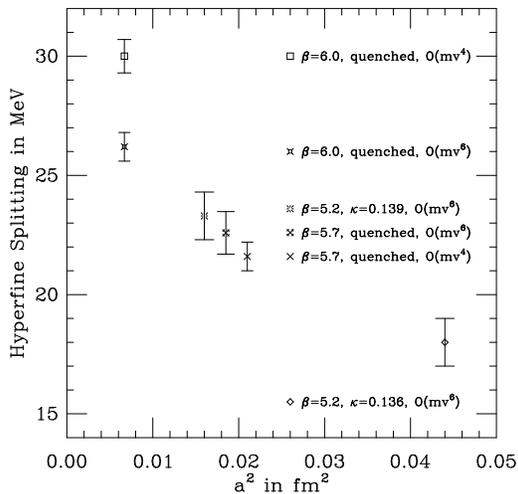}}
\caption{Hyperfine Structure in $\Upsilon$. The values around $a^2 \approx 0.02~\mbox{fm}^2$ are
shifted for clarity.}
\label{fig:hypersplit}
\end{figure}

For $\Upsilon$ the relativistic corrections to the spin splittings are
expected at the 10\% level from power counting arguments. In Figure
\ref{fig:hypersplit} we show our results for the hyperfine structure,
${}^3S_1-{}^1S_0$. We see a clear effect in this very accurate
quantity and the change of 13(3)\% compares well with this expectation.
For charmonium the results are less encouraging and we see 51(2)\%
effects which are beyond the na\"\i ve expectations from NRQCD.
Similarly, there is also a reduction of the fine structure, ${}^3P_J -
\bar P$, away from the experimental values. This is a 34(14)\%
effect and it is still consistent with our expectations. 
But we note that the fine structure ratio is now consistent with
experiment, $R_{fs}=0.56(19)$. 
This indicates that the inclusion of our new terms is important to
reproduce the internal structure of the P-wave triplet accurately.
${}^3P_J - \bar P$. 
\begin{figure}[htb]
    \hbox{\epsfxsize = 72mm  \epsffile{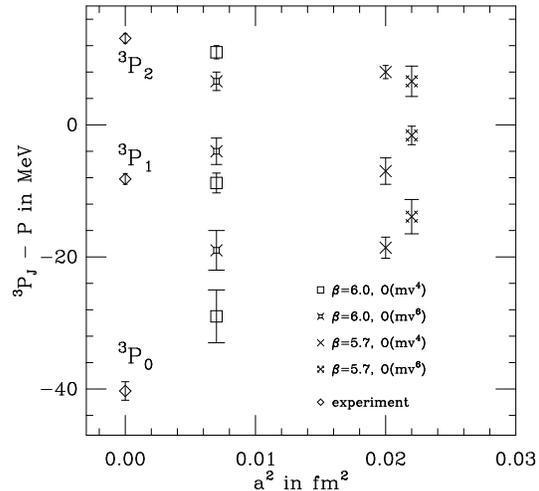}}
\vskip -12pt
\caption{Fine Structure in $\Upsilon$. The scale is
determined from the $1P-1S$ splitting.}
\label{fig:finesplit}
\end{figure}
\subsection{$\beta=5.7$ vs. $\beta=6.0$}
There will be lattice artefacts from discretisation errors which are
not removed at the tree level. In Figure \ref{fig:hypersplit} and
\ref{fig:finesplit} all results are plotted against $a^2$, determined from the
$1P-1S$ splitting. 
We see that our new evolution equation has reduced the lattice spacing
dependence in the spin structure significantly. The remaining effects
can be interpreted as being
$O(\alpha a^2)$ \cite{scaling}. A different tadpole
prescription may reduce these discretisation errors further 
\cite{trottier}. Comparing our results to those in \cite{achim}
we see that going from $u_{0P} \to u_{0L}$ changes the hyperfine
splitting by the same magnitude as the change from $O(mv^4) \to
O(mv^6)$ but in the opposite direction. We expect a similar behaviour
for the P-state triplet.
\subsection{$N_f=0$ vs. $N_f=2$}
We have also used preliminary data from unquenched configurations
at $\beta=5.2$ with 2 dynamical flavours \cite{talevi}. 
Our values with $\kappa_{\rm sea}=0.139$ are directly comparable
to the results at $\beta=5.7, N_f=0$. For this sea quark mass we see a
very weak dependence on the number of flavours. The systematic errors
discussed above have certainly a larger effect. Therefore unquenching effects
should be  studied for spin-independent quantities where
spin corrections cannot obscure the analysis. However,
those quantities have a bigger statistical errors and the observed
effects are small as well \cite{achim}.
\section{HEAVY HYBRIDS}
\begin{figure}[htb]
    \hbox{\epsfxsize = 72mm  \epsffile{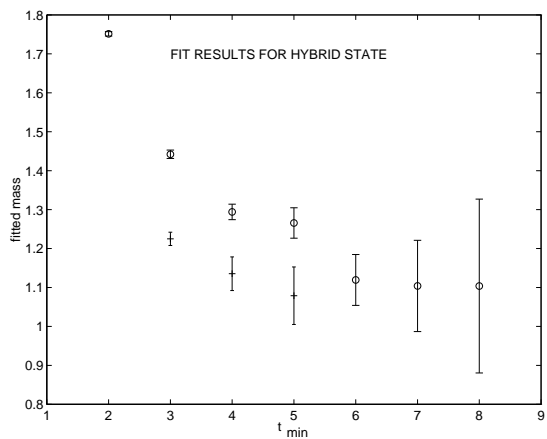}}
\vskip -0.5cm
\caption{$t_{\rm min}$ plot for a one-exponential (circles) and a two-exponential (crosses)
fit. $t_{\rm max}=10$.}
\label{fig:hybridtmin}
\end{figure}

\begin{figure}[htb]
    \hbox{\epsfxsize = 70mm  \epsffile{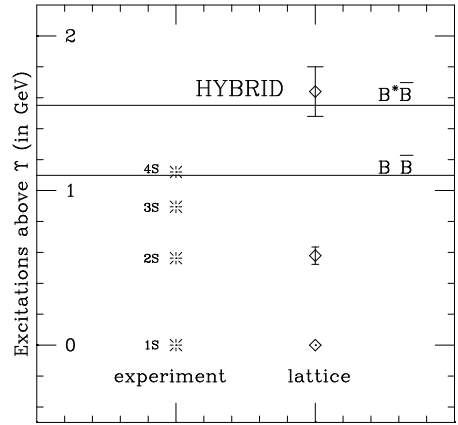}}
\vskip -0.5cm
\caption{Spin-independent excitations in $\Upsilon$. The scale is
determined from the $1P-1S$ splitting.}
\label{fig:hybridexp}
\end{figure}

From phenomenological models one expects the lowest hybrid state
to come from a magnetic gluon $J^P=1^+$ which couples to $Q\bar Q$
in an octet representation. The simplest lattice operators for such a
state are 
\be
H_{ij}^{-+}(x) = \chi^{\dag}(x)\sigma_i B_j \Psi(x)
\ee
and the corresponding spin-singlet without $\sigma_i$. On a lattice they
belong to the representations $A^{-+}, T_1^{-+}, T_2^{-+}, E^{-+}, T_1^{--}$. This list
also includes the exotic hybrid $1^{-+}$.
As a first step in establishing the masses of the lowest lying hybrid
states we employed the leading order NRQCD correct to $O(mv^2)$.
At this level of accuracy all these states are degenerate in energy.
The optimisation of the overlap with the hybrid state is more elaborate 
and requires most of the CPU time.
We use a fuzzing algorithm for the link variables from which 
we construct the extended operators with correct $J^{PC}$.
We calculated propagators for several combinations of operators with
different extent in the sink and source. 
After averaging over 20,000 sources from 500 configurations at
$\beta=6.0$ we obtained a very clear signal. We show the correlated
fit results in Figure \ref{fig:hybridtmin}. 
To illustrate this result we compare the spin independent spectrum
from NRQCD with the experimental values in Figure \ref{fig:hybridexp}. 
The hybrid state is our
prediction and it is approximately 1.64(16) GeV above the $\Upsilon$.
A more careful analysis is needed to study the systematic errors.
In particular the finite volume errors are of concern because such states 
will be rather broad on the lattice we used ($16^3 \times 48,
a^{-1}=2.44(4)$ GeV).
Future work will also include the spin-dependent terms in NRQCD to study
the splittings within the hybrid multiplets.

\end{document}